# RECOVERING THE SHAPE OF A QUANTUM CATERPILLAR TREE BY TWO SPECTRA

## D. Kaliuzhnyi-Verbovetskyi[1], V. Pivovarchik[1]

[1]*South Ukrainian National Pedagogical University named after K. D. Ushynsky*

**Abstract:** It is known that there exist co-spectral (iso-spectral) quantum graphs. In other words, the spectrum of the Sturm-Liouville problem on a metric graph does not determine the shape of the graph.

We consider two Sturm-Liouville spectral problems on an equilateral metric caterpillar tree with real $L_2(0,l)$ potentials on the edges. In the first (Neumann) problem we impose standard conditions at all vertices: Neumann boundary conditions at the pendant vertices and continuity and Kirchhoff's conditions at the interior vertices. The second (Dirichlet) problem differs from the first in that in the second problem we set the Dirichlet condition at the root (one of the pendant vertices of the stalk of the caterpillar tree). Using the asymptotics of the eigenvalues of these two spectra we find the determinant of the normalized Laplacian of the tree and the determinant of the prime submatrix of the normalized Laplacian. Expanding the fraction of these determinants into continued fraction we receive full information on the shape of the tree. We prove that in the case of a caterpillar tree the spectra of the Neumann and Dirichlet problems uniquely determine the shape of the tree.

**Keywords:** metric graph, tree, pendant vertex, interior vertex, edge, caterpillar tree, Sturm-Liouville equation, potential, eigenvalues, spectrum, Dirichlet boundary condition, Neumann boundary condition, root, continued fraction, adjacency matrix, prime submatrix, normalized Laplacian.

## 1 INTRODUCTION

The problem of recovering the shape of a combinatorial graph using the eigenvalues of its adjacency matrix is described in [4] (Chapter 6) where several examples of co-spectral graphs are shown.

In quantum graph theory, i. e. in the theory of quantum mechanical equations considered on metric graph domains, the problem of recovering the shape of a graph was stated in [1] and [8]. It was shown in [8] that if the lengths of the edges are non-commensurate then the spectrum of the spectral Sturm-Liouville problem on a graph with standard (continuity + Kirchhoff's at the interior vertices and the Neumann at the pendant vertices) conditions uniquely determines the shape of this graph.

In [1], it was shown that in case of commensurate lengths of the edges there exist co-spectral quantum graphs. But even earlier it was shown in [2] that in quantum graphs theory an important role is played not by adjacency matrix but by the so-called normalized Laplacian.

A 'geometric' Ambarzumian's theorem was proved in [9]: it was shown that the spectrum of the Neumann problem with zero potential on the graph $P_2$, i.e. on a finite interval, uniquely determines the shape of the graph. In [6] it was shown that if the graph is simple connected equilateral with the number of vertices less or equal 5 and the potentials on the edges are real $L_2$ functions then the spectrum of the Sturm-Liouville problem with standard conditions at the vertices uniquely determines the shape of the graph. For trees the minimal number of vertices in a co-spectral pair is 9 (see [12] and [7]). If the number of vertices doesn't exceed 8 then to find the shape of a tree we need just to find in [6] the characteristic polynomial corresponding to the given spectrum.

In [12] it was shown how to find the shape of a tree using the two spectra: the spectrum of the Neumann problem and the spectrum of the Dirichlet problem, i. e. the problem in which the Dirichlet condition is imposed at the root. This method works even in case of large number of vertices. If the solution is not unique, we can find all the solutions. In [110] it was shown how to find the shape of a tree using the S-function of the scattering problem on a tree which consists of an equilateral compact subtree with a lead attached to it. The potential on the lead was assumed to be zero identically and therefore the Jost-function can be expressed via the characteristic functions of the Dirichlet and Neumann problems. Thus, this scattering inverse problem and the spectral inverse problem by two spectra are closely related.

In present paper we show that in case of a caterpillar tree rooted at a pendant vertex of the stalk (central path) the spectra of the Dirichlet and Neumann problems uniquely determine the shape of the tree.

In Section 2 we describe the Neumann spectral problem, i. e. the Sturm-Liouville problem with standard conditions (continuity + Kirchhoff's at the interior vertices and Neumann at the pendant vertices). We describe the Dirichlet problem where we impose the Dirichlet condition at the root (an arbitrary chosen vertex) keeping standard conditions at all the other vertices. We also we expose known results which we use in the sequel.

In Section 3 we prove a theorem where the fraction of the characteristic polynomial of the normalized Laplacian of a caterpillar combinatorial tree and the modified characteristic polynomial of its certain subgraph obtained by deleting the root and the incident edge is presented as a branched continuous fraction. We prove that in case of a caterpillar tree this presentation is unique.

In Section 4 using the result of Section 3 we show the procedure of recovering the shape of a tree using asymptotics of the spectra of the Neumann and Dirichlet problems.

## 2   STATEMENT OF THE PROBLEM AND AUXILIARY RESULTS

**Definition 2.1**, [5] A combinatorial caterpillar tree is a tree in which all the vertices are within distance 1 of a central path (stalk).

An example of a caterpillar tree is presented on Fig. 1.

Let $T$ be a metric equilateral caterpillar graph with $p$ vertices and $g = p-1$ edges each of the length $l$. Let $v_0 \to v_1 \to \ldots \to v_r$ be the stalk of the graph (the longest path). It means that the degree $d(v_0) = d(v_r) = 1$.

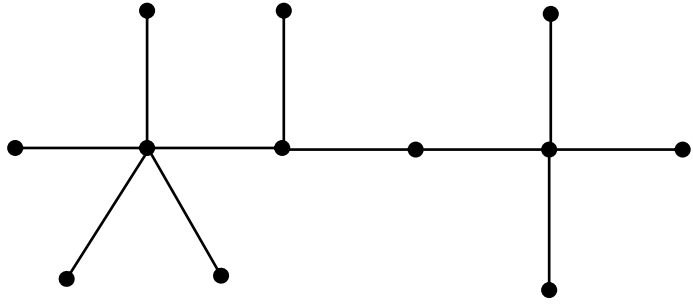

**Fig. 1. An example of caterpillar tree.**

We choose the vertex $v_0$ as the root and direct all the edges away from the root.

Let us describe the Neumann spectral problem on this tree. We consider the Sturm-Liouville equations on the edges

$$-y_j'' + q_j(x) y_j = \lambda y_j, \quad j = 1, 2, \ldots, g \tag{1}$$

where $q_j \in L_2(0, l)$ are real.

If an edge $e_j$ is incident with a pendant vertex which is not the root then we impose the Neumann condition

$$y_j'(l) = 0. \tag{2}$$

at the pendant vertex. At each interior vertex we impose the continuity conditions

$$y_j(l) = y_k(0) \tag{3}$$

for the incoming into $v_i$ edge $e_j$ and for all $e_k$ outgoing from $v_i$, and the Kirchhoff's conditions

$$y_j'(l) = \sum_k y_k'(0) \tag{4}$$

where the sum is taken over all edges $e_k$ outgoing from $v_i$. At the root we impose the Neumann condition:

$$y_1'(0) = 0. \tag{5}$$

The above conditions (continuity +Kirchhoff's or Neumann) we call standard.

In the sequel, if the potentials are the same on all the edges, we omit the index in $q_j$ and $y_j$. The following theorem adopted for trees can be found as Theorem 5.2 in [6] but it originates from [2].

**Theorem 2.1** *Let T be a tree with $p \geq 2$. Assume that all edges have the same length l and the same potentials symmetric with respect to the midpoints of the edges $(q(l-x) = q(x))$. Then the spectrum of problem (1)–(5) coincides with the set of zeros of the function*

$$\varphi_N(\lambda) = s(\sqrt{\lambda}, l) \omega(c(\sqrt{\lambda}, l))$$

*where* $\omega(z) = (1-z^2)^{-1} \psi(z),$

$$\psi(z) = \det(-zD + A).$$

*Here A is the adjacency matrix of T in which the first row and the first column correspond*

to $v_0$,

$$D := diag(d(v_0), d(v_1), ..., d(v_{p-1})),$$

$d(v_i)$ is the degree of the vertex $v_i$, $s(\sqrt{\lambda}, x)$ and $c(\sqrt{\lambda}, l)$ are the solutions of the Sturm-Liouville equation on the edges satisfying conditions $s(\sqrt{\lambda}, 0) = s'(\sqrt{\lambda}, 0) - 1 = 0$ and $c(\sqrt{\lambda}, 0) - 1 = c'(\sqrt{\lambda}, 0)$.

Now we consider the Dirichlet problem on the same caterpillar tree. We impose the Dirichlet condition at $v_0$:

$$y_1(0) = 0 \qquad (6)$$

for the edge incident with $v_0$.

By the Dirichlet problem we mean the problem which consists of equations (1)–(4) and (6).

Denote by $T_1$ the tree obtained by removing the root in the tree $T$ together with the incident edge. Let $A_1$ be the adjacency matrix of $T_1$, i. e. the principal submatrix of $A$ obtained by deleting the first row and the first column of $A$, let $D_1$ be the principal submatrix of $D$ obtained by deleting the first row and the first column of $D$.

We consider the polynomial defined by

$$\theta(z) := \det(-zD_1 + A_1).$$

Theorem 6.4.2 of [10] adapted to the case a tree with the Dirichlet condition at one of the vertices is as follows

**Theorem 2.2** *Let $T$ be a tree with at least two edges rooted at a pendant vertex $v_0$. Let the Dirichlet condition be imposed at the root and the standard conditions at all other vertices. Assume that all edges have the same length l and the same potentials symmetric with respect to the midpoints of the edges $(q(l-x) = q(x))$. Then the spectrum of problem (1)–(4), (6) coincides with the set of zeros of the characteristic function*

$$\varphi_D(\lambda) = \theta(c(\sqrt{\lambda}, l)).$$

It is clear that

$$\varphi_D(\lambda) = \det(-c(\sqrt{\lambda}, l)D_1 + A_1)$$

is the characteristic function of the Dirichlet problem (1) - (4), (6) on the initial tree T.

## 3 MAIN RESULTS

The following theorem was proved in [12].

**Theorem 3.1.** Let be an equilateral tree. Then the function $\dfrac{\psi(z)}{\theta(z)}$ can be presented as a branched continued fraction. The coefficients before +z and -z correspond to the degrees of the vertices. The beginning fragment

$$-m_0 z + \sum_{k=1}^{m_0} \frac{1}{m_k z - \ldots}$$

of the expansion means that the vertex $v_0$ is connected by edges with $m_0$ vertices $v_1, v_2, \ldots, v_{m_0}$.
A fragment

$$\ldots \pm \sum_{i=1}^{r} \frac{1}{-m_i z + \sum_{k=1}^{m_i - 1} \frac{1}{m_{i,k} z - \ldots}}$$

means that there are there are r vertices of degrees $m_1, m_2, \ldots, m_r$ each having one incoming edge and $m_1 - 1, m_2 - 1, \ldots, m_r - 1$ outgoing edges.
A fragment

$$\ldots \pm \frac{m}{z}$$

at the end of a branch of the continued fraction means m edges ending at pendant vertices.

This theorem applied to a caterpillar tree rooted and one of its pendant vertices gives

**Corollary 3.2.** Let T be a caterpillar tree rooted at $v_0$, one of the ends of the stalk. Then the fraction $\frac{\psi(z)}{\theta(z)}$ can be presented as

$$\frac{\psi(z)}{\theta(z)} = -z + \cfrac{1}{m_1 z - \cfrac{m_1 - 2}{z} - \cfrac{1}{m_2 z - \cfrac{m_2 - 2}{z} - \ldots - \cfrac{1}{m_{r-1} z - \cfrac{m_{r-1} - 1}{z}}}} \qquad (7)$$

where $\{1, m_1, m_2, \ldots, m_{r-1}, 1\}$ are the degrees of stalk vertices.

**Theorem 3.3.** If the fraction $\frac{\psi(z)}{\theta(z)}$ can be expanded into continued fraction of the form (7) with integers $m_i \geq 2$ for i=1,2, …, r-1 then there exists a unique tree which is a caterpillar tree rooted at $v_0$ with the degrees of the vertices on the stalk $d(v_0) = d(v_r) = 1$ and $d(v_i) = m_i$ for i = 1, 2, …, r-1.

**Proof**. The coefficient $m_1$ is uniquely determined as

$$m_1 = \lim_{z \to \infty} \left( \frac{z \psi(z)}{\theta(z)} + z^2 \right)^{-1}.$$

Then the coefficient $m_2$ is uniquely determined as

$$m_2 = \lim_{z \to \infty} \left( -\left( \left( \frac{z \psi(z)}{\theta(z)} + z^2 \right)^{-1} - m_1 \right) z^2 - m_1 + 2 \right)^{-1}$$

we continue this procedure and obtain all $m_i$. On each stage of this procedure, we face the Diophantine equation

$$\sum_{k=1}^{m_i-1} \frac{1}{n_k} = m_i - 2 + \frac{1}{m_{i+1}}$$

with respect to integer unknowns $n_k \geq 1$. Since $m_i - 2 < m_i - 2 + \frac{1}{m_{i+1}} \leq m_i - 1$ we conclude that equation possesses a unique up to permutations solution $n_1 = n_2 = ... = n_{m_i-2} = 1$, $n_{m_i-1} = m_{i+1}$.
QED.

**Example**. Let $\psi(z) = -120z^7 + 269z^5 - 189z^3 + 40z$ and $\theta(z) = 120z^6 - 245z^4 + 156z^2 - 30$. Then

$$\frac{\psi(z)}{\theta(z)} = -z + \frac{24z^5 - 33z^3 + 10z}{120z^6 - 245z^4 + 156z^2 - 30} = -z + \cfrac{1}{5z + \cfrac{-80z^4 + 106z^2 - 30}{24z^5 - 33z^3 + 10z}}.$$

Since $3 < \frac{80}{24} < 4$, we present the fraction as

$$\frac{\psi(z)}{\theta(z)} = -z + \cfrac{1}{5z - \cfrac{3}{z} - \cfrac{8z^3 - 7z}{23z^4 - 33z^2 + 10}} = -z + \cfrac{1}{5z - \cfrac{3}{z} - \cfrac{1}{3z - \cfrac{12z^2 - 10}{8z^3 - 7z}}}.$$

Since $1 < \frac{12}{8} < 2$ we arrive at

$$\frac{\psi(z)}{\theta(z)} = -z + \cfrac{1}{5z - \cfrac{3}{z} - \cfrac{1}{3z - \cfrac{1}{z} - \cfrac{4z^2 - 3}{8z^3 - 7z}}} = -z + \cfrac{1}{5z - \cfrac{3}{z} - \cfrac{1}{3z - \cfrac{1}{z} - \cfrac{1}{2z - \cfrac{z}{4z^2 - 3}}}}$$

and finally

$$\frac{\psi(z)}{\theta(z)} = -z + \cfrac{1}{5z - \cfrac{3}{z} - \cfrac{1}{3z - \cfrac{1}{z} - \cfrac{1}{2z - \cfrac{1}{4z - \cfrac{3}{z}}}}}.$$

Judging by this continued fraction we conclude that if $\psi(z) = -120z^7 + 269z^5 - 189z^3 + 40z$ and $\theta(z) = 120z^6 - 245z^4 + 156z^2 - 30$ then the corresponding tree is the that of Fig.1.

## 4 RECOVERING THE SHAPE OF A QUANTUM TREE BY TWO SPECTRA

Now we are ready to recover the shape of a caterpillar tree. Using the asymptotics of the spectrum of the Neumann problem we can find the function $\psi(z)$ (up to a constant factor). Let us show it.

By Theorem 2.1 in case of $q_j(x) \equiv 0$ for all j, the spectrum of problem (1)-(5) can be presented as the union of subsequences $\{\lambda_{k,0}\}_{k=1}^{\infty} = \bigcup_{i=1}^{2g-1} \{\lambda_{k,0}^{(i)}\}_{k=1}^{\infty}$ with the following asymptotics

$$\sqrt{\lambda_{k,0}^{(i)}} \underset{k\to\infty}{=} \frac{2\pi(k-1)}{l} + \frac{1}{l}\arccos\alpha_i \quad \text{for } i = 2, 3, \ldots, p\text{-}1, \quad k \in \mathbb{N},$$

$$\sqrt{\lambda_{k,0}^{(i)}} \underset{k\to\infty}{=} \frac{2\pi k}{l} - \frac{1}{l}\arccos\alpha_{-p+2+i} \quad \text{for } i = p, p+1, \ldots, 2p\text{-}3, k \in \mathbb{N},$$

$$\sqrt{\lambda_{k,0}^{(1)}} \underset{k\to\infty}{=} \frac{\pi(k-1)}{l} \quad \text{for } k \in \mathbb{N},$$

where $\alpha_1 = 1 \leq \alpha_2 \leq \ldots \leq \alpha_{p-1} \leq \alpha_p = 1$ are the zeros of $\psi(z)$.

By Theorem 5.4 in [3] we obtain that there exists a positive constant C such that $|\lambda_k - \lambda_{k,0}| < C$ where $\lambda_k$ are eigenvalues of problem (1)-(5) with $L_2(0,l)$ potentials on the edges.

**Theorem 4.1.** Let T be an equilateral caterpillar tree with p vertices and with real potentials $q_j(x) \in L_2(0,l)$ on the edges. Then the spectrum of problem (1)-(5) can be presented as the union of subsequences $\{\lambda_k\}_{k=1}^{\infty} = \bigcup_{i=1}^{2g-1}\{\lambda_k^{(i)}\}_{k=1}^{\infty}$ with the following asymptotics

$$\sqrt{\lambda_k^{(i)}} \underset{k\to\infty}{=} \frac{2\pi(k-1)}{l} + \frac{1}{l}\arccos\alpha_i + O\left(\frac{1}{k}\right) \quad \text{for } i = 2, 3, \ldots, p\text{-}1, \quad k \in \mathbb{N}$$

$$\sqrt{\lambda_k^{(i)}} \underset{k\to\infty}{=} \frac{2\pi k}{l} - \frac{1}{l}\arccos\alpha_{-p+2+i} + O\left(\frac{1}{k}\right) \quad \text{for } i = p, p+1, \ldots, 2p\text{-}3, \quad k \in \mathbb{N}$$

$$\sqrt{\lambda_k^{(1)}} \underset{k\to\infty}{=} \frac{\pi(k-1)}{l} + O\left(\frac{1}{k}\right) \quad \text{for } k \in \mathbb{N},$$

where $\alpha_1 = 1 \leq \alpha_2 \leq \ldots \leq \alpha_{p-1} \leq \alpha_p = 1$ are the zeros of $\psi(z)$.

By Theorem 2.2 in case of $q_j(x) \equiv 0$ for all j, the eigenvalues of problem (1)-(4), (6) can be presented as the union of subsequences $\{\nu_{k,0}\}_{k=1}^{\infty} = \bigcup_{i=1}^{2g}\{\nu_{k,0}^{(i)}\}_{k=1}^{\infty}$ with the following asymptotics

$$\sqrt{\nu_{k,0}^{(i)}} \underset{k\to\infty}{=} \frac{2\pi(k-1)}{l} + \frac{1}{l}\arccos\beta_i \quad i=1, 2, \ldots, p\text{-}1, k \in \mathbb{N},$$

$$\sqrt{\nu_{k,0}^{(i)}} \underset{k\to\infty}{=} \frac{2\pi k}{l} - \frac{1}{l}\arccos\beta_{-p+1+i} \quad i=p, p+1, \ldots, 2p\text{-}2, k \in \mathbb{N}.$$

Again using Theorem 5.4 in [3] we obtain that there exists a positive constant C such that $|v_k - v_{k,0}| < C$ where $\lambda_k$ are eigenvalues of problem (1)-(4), (6) with $L_2(0,l)$ potentials on the edges.

**Theorem 4.2.** Let T be an equilateral caterpillar tree with p vertices and with real potentials $q_j(x) \in L_2(0,l)$ on the edges. Then the spectrum of problem (1)-(4), (6) can be presented as the union of subsequences $\{v_k\}_{k=1}^{\infty} = \bigcup_{i=1}^{2g-1} \{v_k^{(i)}\}_{k=1}^{\infty}$ with the following asymptotics

$$\sqrt{v_k^{(i)}} \underset{k \to \infty}{=} \frac{2\pi(k-1)}{l} + \frac{1}{l}\arccos\beta_i + O\left(\frac{1}{k}\right) \quad i=1, 2, \ldots, p-1, k \in \mathbb{N}$$

$$\sqrt{v_k^{(i)}} \underset{k \to \infty}{=} \frac{2\pi k}{l} - \frac{1}{l}\arccos\beta_{-p+1+i} + O\left(\frac{1}{k}\right) \quad i=p, p+1, \ldots, 2p-2, k \in \mathbb{N},$$

where $\{\beta_i\}_{k=1}^{p-1}$ are the zeros of $\theta(z)$.

According to Theorems 4.1 and 4.2 using the two spectra $\{\lambda_k\}_{k=1}^{\infty}, \{v_k\}_{k=1}^{\infty}$ we can find the sets of zeros of the numerator and denominator of the rational function $\frac{\psi(z)}{\theta(z)}$. Thus, this function is uniquely determined if we take into account that (7) implies

$$\lim_{z \to \infty} \frac{\psi(z)}{z\theta(z)} = -1.$$

Expanding $\frac{\psi(z)}{\theta(z)}$ into continued fraction (7) we find the shape of our caterpillar tree.

## 5 CONCLUSIONS

As it was mentioned in the introduction that in general the spectrum of a Sturm-Liouville spectral problem on a simple connected equilateral graph does not determine uniquely the shape of the graph. We don't know whether two spectra of such problems with different conditions on the same graph uniquely determine the shape of the graph. However, we describe a class of trees (caterpillar trees) for which the two spectra uniquely determine the shape of a graph. We also give an algorithm of recovering the shape of a caterpillar tree.

**Acknowledgements**
The second author is grateful to the Ministry of Education and Science of Ukraine for the support in completing the work 'Artificial porous materials as a basis for creating the novel biosensors'.

**Kaliuzhnyi-Verbovetskyi Dmytro**
South Ukrainian National Pedagogical University named after K.D. Ushynsky,
Doctor of Physical and Mathematical Sciences,
Staroportofrankovskaya 26, Odesa, Ukraine 65020
dmitry2k@yahoo.com
ORCID: 0000-0002-7411-3740

**Pivovarchik Vyacheslav**
South Ukrainian National Pedagogical University named after K.D. Ushynsky,
Doctor of Physical and Mathematical Sciences,
Staroportofrankovskaya 26, Odesa, Ukraine 65020
vpivovarchik@gmail.com,
ORCID: 0000-0002-4649-2333